\documentclass[preprint,aps,amsmath,nofootinbib,english,12pt]{revtex4}
\usepackage{graphics,color,array,dcolumn}
\usepackage{graphicx}
\usepackage{amssymb}
\usepackage{xspace}
\usepackage{color}

\makeatother
\begin{document}

\title{Fixed Points of Nonlinear Sigma Models in $d>2$}
\bigskip
\author{Alessandro Codello}
\email{a.codello@gmail.com}
\affiliation{Institut f\"ur Physik, Johannes-Gutenberg-Universit\"at, Staudingerweg 7, D-59099 Mainz, Germany}
\author{Roberto Percacci}
\email{percacci@sissa.it}
\affiliation{SISSA, via Beirut 4, I-34014 Trieste, Italy, and INFN, Sezione di Trieste, Italy}
\medskip
\begin{abstract}
Using Wilsonian methods, we study the renormalization group flow
of the Nonlinear Sigma Model in any dimension $d$,
restricting our attention to terms with two derivatives.
At one loop we always find a Ricci flow.
When symmetries completely fix the internal metric, we compute the beta function 
of the single remaining coupling, without any further approximation.
For $d>2$ and positive curvature, 
there is a nontrivial fixed point, which could be used to define 
an ultraviolet limit, in spite of the
perturbative nonrenormalizability of the theory.
Potential applications are briefly mentioned.
\end{abstract} 

\maketitle

The Nonlinear Sigma Models (NLSMs) are a very rich class of theories,
describing the dynamics of a map $\varphi$ from a $d$-dimensional 
manifold $M$ to a $D$-dimensional manifold $N$.
They have been applied to phenomenological models of high energy physics,
to condensed matter systems, as well as strings and branes
\footnote{in string and brane theories, spacetime is identified with $N$.
Here we stick to the field-theoretic interpretation
where spacetime is identified with $M$. Since we are not interested
in gravity, $M$ is flat.} \cite{Ketov}.
Given coordinate systems $\{x^\mu\}$ on $M$ and $\{y^\alpha\}$ on $N$,
one can describe the map $\varphi$ by $D$ scalar fields $\varphi^\alpha(x)$.
Physics must be independent of the choice of coordinates on $N$,
forcing the action to be a functional constructed with tensorial
structures on $N$. Only derivative interactions are allowed.
Linear scalar theories correspond to the case
when $N$ is a linear space. In this case (and only in this case)
one can chose the action to describe free fields,
and interactions are usually provided by a potential.
Thus the NLSMs are profoundly different from linear scalar theories.

The action of the NLSM can be expanded in derivatives and the lowest term is:
\begin{equation}
\label{action}
\frac{1}{2}\zeta\int d^dx\,\partial_{\mu}\varphi^\alpha\partial^\mu\varphi^\beta h_{\alpha\beta}(\varphi)\ .
\end{equation}
where $h_{\alpha\beta}$ is a dimensionless metric and $\zeta=1/g^2$ has dimensions mass$^{d-2}$.
Applying the formalism of quantum field theory to these models
requires some adaptation.
The simplest treatment is based on the assumption that the ground state
of the theory is a constant map $\bar\varphi$.
There exists a local diffeomorphism $Exp_{\bar\varphi}$
of the tangent space $T_{\bar\varphi}N$
to a neighborhood $U$ of $\bar\varphi$, given by mapping a vector $\xi$ 
to the point lying a distance $||\xi||$ along the geodesic emanating
from $\bar\varphi$ in the direction of $\xi$.
The components $\xi^\alpha$ can be used as coordinates on $U$,
called normal coordinates.
Fluctuations around the vacuum are faithfully described
by the fields $\xi^\alpha(x)$, which can be quantized by path integral methods.
When the action is thus expanded around $\bar\varphi$ 
and the fields are canonically normalized, one recognizes
that $g$ plays the role of coupling constant, and since it has dimension
of mass$^{\frac{2-d}{2}}$, this perturbative expansion is nonrenormalizable
for $d>2$.
As a consequence, phenomenological applications of the NLSM in $d$=4 are regarded
as effective field theories with a cutoff.

Here we are interested in the possibility that some of these theories in $d>2$
may actually be nonperturbatively renormalizable, in the sense that
the continuum limit can be taken at a nontrivial Fixed Point (FP) of
the Renormalization Group (RG). To establish this property one should
in principle compute the beta functions of all possible couplings.
If they admit a FP with a finite number of UV--attractive (relevant) directions, 
then the theory is ``asymptotically safe'' \cite{Weinberg}:
it has a sensible UV limit and is predictive, because only the
relevant couplings need to be fixed from experiment.
We provide here some new evidence that certain NLSMs in $d>2$, 
including $d=4$, may have these properties.
For previous work in $2+\epsilon$ dimensions see \cite{polyakov,friedan,ZJ,BLS}
and in three dimensions see \cite{arefeva,itou}.

We shall begin by evaluating the
beta functions in the one loop approximation.
We use the background field method, expanding
$\varphi^\alpha(x)=\bar\varphi^\alpha(x)+\eta^\alpha(x)$.
For reasons that will become clear soon, it will not be sufficient to
consider constant backgrounds, so the
simple procedure described above will have to be generalized.
Furthermore, the field $\eta$ is a difference of coordinates and
does not have good transformation properties.
The treatment of the NLSM with general backgrounds has been
discussed by several authors \cite{honerkamp,AFM,HPS}.
Basically, at each point $x\in M$
one evaluates the Lagrangian density $L(x)$ using the
normal coordinates centered at $\bar\varphi(x)$.
They are the components of a section $\xi$ of $\bar\varphi^*TN$,
such that $Exp_{\bar\varphi(x)}(\xi(x))=\varphi(x)$.
The field $\eta$ can be written as a function of $\xi$,
which is taken as the quantum field.
One can expand $L(x)$ in $\xi$ and write the result in a tensorial form,
in such a way that invariance under background coordinate transformations
is manifest.

We study the RG in a ``Wilsonian'' fashion, introducing by hand 
an infrared cutoff $k$ in the theory and calculating the dependence of the
effective action on $k$.
The cutoff is a term quadratic in the quantum fields $\xi$, of the form
$\Delta S_k(\bar\varphi,\xi)=\frac{1}{2}\int dx\,\xi^\alpha ({\cal R}_k)_{\alpha\beta}\xi^\beta$.
The kernel ${\cal R}$, to be specified later,
is chosen in such a way that
it suppresses the propagation of the modes with momenta $q^2<k^2$,
leaving the modes with momenta $q^2>k^2$ unaffected.
In the limit of an infinitely strong suppression this is equivalent
to a sharp IR cutoff on the path integration. 
The generating functional of connected Green functions $W_k(\bar\varphi,j)$
is defined by 
$$
e^{-W_k[\bar\varphi,j]}=\int (d\xi) \exp\left(-S[\varphi]-\Delta S_k[\bar\varphi,\xi]-\int j\cdot\xi\right)
$$
and the $k$--dependent effective action is given by the modified Legendre transform 
\cite{Wetterich}
$$
\bar\Gamma_k[\bar\varphi,\xi]=W_k[\bar\varphi,j(\xi)]-\int j\cdot \xi -\Delta S_k[\bar\varphi,\xi]\ .
$$
Taken at tree level, it describes the effective dynamics at the energy scale $k$.
We will be especially interested in the functional
$\Gamma_k(\bar\varphi)=\bar\Gamma_k(\bar\varphi,0)$.
At one loop it is given by
$$
\Gamma_k^{(1)}(\bar\varphi)=
S(\bar\varphi)
+\frac{1}{2}\mathrm{Tr}\log\frac{\delta^2 (S+\Delta S_k)}{\delta\xi\delta\xi}\Bigg|_{\bar\varphi}\ .
$$
Its logarithmic derivative with respect to $k$ is
the one loop ``beta functional''
\begin{equation}
\label{onelooperge}
\dot\Gamma_k^{(1)}(\bar\varphi)=
\frac{1}{2}\mathrm{Tr}\left(\frac{\delta^2 S}{\delta\xi\delta\xi}+{\cal R}_k\right)^{-1}
\dot{\cal R}_k\ .
\end{equation}
Here an overdot denotes derivative with respect to $t=\log(k/k_0)$.
In order to calculate the beta function of the metric $\zeta h_{\alpha\beta}$
we have to extract from the trace on the r.h.s. the term containing
two derivatives.
It is convenient to define the quantum fields $\xi^a=e^a_\alpha\xi^\alpha$,
where $e^a_\alpha$ is a vielbein for the metric $h_{\alpha\beta}$.
Then, the quadratic part of the action (1) is \cite{AFM,HPS}
\begin{equation}
\label{quadratic}
\frac{1}{2}\zeta\int dx\,\xi^{a}\left(-D^{2}\delta_{ab}-M_{ab}\right)\xi^{b}\ ,
\end{equation}
where $D_\mu\xi^\alpha=\partial_\mu\xi^\alpha+
\partial_\mu\bar\varphi^\beta\Gamma_\beta{}^\alpha{}_\gamma\xi^\gamma$,
and
$M_{ab}(\bar\varphi)=
e_{a}^\alpha e_{b}^\beta\partial_\mu\bar\varphi^\gamma\partial^\mu\bar\varphi^\delta 
R_{\alpha\gamma\beta\delta}$.
Here $\Gamma_\beta{}^\alpha{}_\gamma$ and $R_{\alpha\gamma\beta\delta}$ 
are the connection and curvature of the 
metric $h_{\alpha\beta}$, evaluated on the background field $\bar\varphi$.

It is convenient to choose a cutoff kernel of the form
${\cal R}_{kab}=\zeta\delta_{ab}R_k(-D^{2})$,
where $R_k(z)$ is a function that goes to zero for $z>k^2$ and
to $k^2$ for $z\to 0$. 
In this way the modified inverse propagator is $\zeta(P_k(-D^{2})\delta_{ab}-M_{ab})$,
where $P_k(z)=z+R_k(z)$.
Introducing in (\ref{onelooperge}) and expanding 
in the matrix $\mathbf{M}\equiv\{M_{ab}\}$, we have
\begin{equation}
\label{expansion}
\dot\Gamma_k^{(1)}(\bar\varphi)=\frac{1}{2}\textrm{Tr}\,\frac{\dot{R}_{k}\mathbf{1}}{P_{k}\mathbf{1}-\mathbf{M}}
= \frac{1}{2}\textrm{Tr}\left(\frac{\dot{R}_{k}}{P_{k}}\mathbf{1}
+\frac{\dot{R}_{k}}{P_{k}^2}\mathbf{M}
+\frac{\dot{R}_{k}}{P_{k}^3}\mathbf{M}^2
+O(\mathbf{M}^{3})\right)\ .
\end{equation}
Note that the ``bare'' $\zeta$ is $k$--independent and therefore cancels out
between numerator and denominator.
The term with two derivatives is the second one.
Using an ``optimized'' cutoff of the form $R_k(z)=(k^2-z)\theta(k^2-z)$ \cite{Litim},
it can be evaluated using methods described in Appendix A of \cite{fofrall}:
\begin{equation}
\frac{1}{2}\textrm{Tr}\,\frac{\dot{R}_{k}}{P_{k}^{2}}\mathbf{M} 
=\frac{1}{2(4\pi)^{d/2}}Q_{\frac{d}{2}}\left(\frac{\dot{R}_{k}}{P_{k}^{2}}\right)\int dx\,\textrm{tr}\,\mathbf{M}
=c_{d}k^{d-2}\int dx\sqrt{g}\, 
\partial_\mu\varphi^\alpha\partial^\mu\varphi^\beta R_{\alpha\beta}\ ,\notag
\end{equation}
where 
$Q_n\left[W\right]=\frac{1}{\Gamma(n)}\int dz\,z^{n-1}W(z)$
and
$c_{d}=\frac{1}{(4\pi)^{d/2}\Gamma(d/2+1)}$.
We assume that the renormalized running effective action $\Gamma_k$ has again
the form (\ref{action})
\footnote{Here we use the same notation for bare and renormalized quantities,
hoping that no confusion arises. One has to remember that in this approach
the renormalized quantities run, not the bare ones.}.
Therefore
$$
\dot{\Gamma}_{k}=\frac{1}{2}\int dx\,
\partial_\mu\varphi^\alpha\partial^\mu\varphi^\beta\frac{d}{dt}(\zeta h_{\alpha\beta}(\varphi))+\ldots
$$
and comparing we obtain the Ricci flow
\begin{equation}
\label{ricciflow}
\frac{d}{dt}(\zeta h_{\alpha\beta}(\varphi))=2c_{d}k^{d-2} R_{\alpha\beta}\,.
\end{equation}
This agrees with \cite{friedan} when $d=2+\epsilon$.

Let us now suppose that the metric $h_{\alpha\beta}$ has some Killing vectors,
generating a Lie group $G$.
Since the cutoff is defined by means of the $G$--invariant
Laplacian $-D^2$, it preserves the $G$ invariance.
Therefore if the initial point of the flow is an invariant metric,
the flow takes place within the restricted class of invariant metrics.
From now on we shall restrict ourselves to homogeneous spaces $N=G/H$ 
admitting a single invariant Einstein metric $h_{\alpha\beta}$, up to scalings.
In this case in equation (\ref{ricciflow}) it is convenient 
to think of $h_{\alpha\beta}$ as being fixed, 
and we interpret the RG flow as affecting only $\zeta$.
The Ricci tensor of $h_{\alpha\beta}$ is $R_{\alpha\beta}=\frac{R}{D}h_{\alpha\beta}$,
where $R$ is the Ricci scalar, therefore
\begin{equation}
\label{betaz1}
\dot\zeta=2c_d \frac{R}{D}k^{d-2}\ .
\end{equation}
The one loop beta function $\beta=\dot{\tilde{g}}$ for the dimensionless coupling
$\tilde g=k^{\frac{d-2}{2}}g$ is
\begin{equation}
\label{oneloopbeta}
\beta=\frac{d-2}{2}\tilde{g}-c_{d}\frac{R}{D}\tilde{g}^{3}\ .
\end{equation}
If $d>2$ and $R>0$ there is a nontrivial FP at
$\tilde g_*^2=\frac{d-2}{2}\frac{D}{c_d R}$.
For large $R$ it occurs at small coupling, where perturbation theory is reliable.
The derivative of the beta function at the FP is $\beta'_*=2-d<0$,
so this FP is UV attractive,
and the mass critical exponent is $\nu=-1/{\beta'_*}=1/(d-2)$ in this 
approximation.
In particular for $N=S^D$, $R=D(D-1)$ and we reproduce the results of
the $2+\epsilon$ expansion for the $SO(D+1)$ model \cite{polyakov,ZJ,BLS}.

Every manifold can be isometrically embedded in a linear space of sufficiently
high dimension, and it is sometimes convenient to regard the NLSM 
as a constrained linear theory.
For example, in the $SO(D+1)$ model, one can start from a linear theory with action
$$
\int d^dx\,\left[
\frac{1}{2}Z\sum_{a=1}^{D+1}\partial_{\mu}\phi^a\partial^\mu\phi^a 
+\frac{1}{2}\lambda(\rho-\bar\rho)^2\right]\ ,
$$
where $\rho=\frac{1}{2}\sum_{a=1}^{D+1}\phi^a\phi^a$ and $Z$, $\lambda$, $\bar\rho$ are
running couplings.
The action (1) can be obtained in the limit $\lambda\to\infty$, 
with the identification $\zeta=2Z\bar\rho$.
It is therefore of some interest to derive the beta function
of the NLSM from the one of the linear theory.
The beta functions of $Z$, $\lambda$ and $\bar\rho$ are given 
e.g. in \cite{Berges}, where the notation 
$\kappa\equiv Z\bar\rho k^{d-2}=\frac{1}{2}\zeta k^{d-2}$ is used.
Evaluating these beta functions with the optimized cutoff 
and taking the limit $\lambda\to\infty$,
the anomalous dimension $\eta_Z\equiv\dot Z/Z\to c_d/\kappa$,
whereas $\dot\kappa\to(2-d-\eta_Z)\kappa+Dc_d=(2-d)\kappa+(D-1)c_d$,
in complete accordance with (\ref{betaz1}).
Since the beta function (\ref{betaz1}) implies a (power law) divergence
for $k\to\infty$,
this means that the divergence is the same in the NLSM and in the 
$\lambda\to\infty$ limit of the linear theory,
in agreement with \cite{Chan}.

As a further check we can compute also the effect of $\tilde g$ 
on the running of the four derivative terms.
There are two such contributions: one comes from 
the $B_4(-D^2)$ heat kernel coefficient in the expansion of 
the first term in (\ref{expansion}), the other from the third term.
We find
\begin{eqnarray}
\dot\Gamma_k&\sim &\frac{1}{2(4\pi)^{d/2}}\int dx\sqrt{g}\, 
\left[
Q_{\frac{d}{2}-2}\left(\frac{\dot{R}_k}{P_k}\right)b_4(-D^2)
+Q_{\frac{d}{2}}\left(\frac{\dot{R}_k}{P_k^3}\right)\textrm{tr}\,\mathbf{M}^2
\right]\notag\\
& &=c_{d}\int dx\sqrt{g}\, 
\partial_\mu\varphi^\alpha\partial^\mu\varphi^\beta\partial_\nu\varphi^\gamma\partial^\nu\varphi^\delta 
\left[\frac{d(d-2)}{4}
\frac{1}{6}R_{\alpha\beta\varepsilon\eta}R_{\gamma\delta}{}^{\eta\varepsilon}
+R_{\alpha\varepsilon\beta\eta}R_\gamma{}^\eta{}_\delta{}^\varepsilon
\right]\ .\notag
\end{eqnarray}
In the case of the $SO(4)$ model in four dimensions ($d=4$, $D=3$, $N=S^3$) the allowed 
four derivative terms in the Lagrangian are
$$
(\ell_1 h_{\alpha\beta}h_{\gamma\delta}+\ell_2 h_{\alpha\gamma}h_{\beta\delta})
\partial_\mu\varphi^\alpha\partial^\mu\varphi^\beta\partial_\nu\varphi^\gamma\partial^\nu\varphi^\delta\ .
$$
The Riemann tensor is 
$R_{\alpha\beta\varepsilon\eta}=h_{\alpha\varepsilon}h_{\beta\eta}-h_{\alpha\eta}h_{\beta\varepsilon}$
and one obtains the beta functions
$$
\dot\ell_1=\frac{2}{3}c_4\ ,\qquad
\dot\ell_2=\frac{4}{3}c_4\ .
$$
When one solves for $\ell_1(k)$ and $\ell_2(k)$, the results diverge logarithmically 
for $k\to\infty$; using the identification $\log k^2=\frac{1}{d-4}$,
the coefficients of the divergence agree with the dimensionally regulated 
one loop calculation in \cite{GL}.

Having checked that this formalism reproduces known results at one loop,
we now go beyond this approximation using Wetterich's equation \cite{Wetterich} 
\begin{equation}
\label{ERGE}
\dot\Gamma_k=
\frac{1}{2}{\rm Tr}\left(\frac{\delta^2\Gamma_k}{\delta\xi\delta\xi}+{\cal R}_k\right)^{-1}\dot {\cal R}_k\ .
\end{equation}
This functional RG equation is very similar to (\ref{onelooperge}),
but it is an exact equation.
Note that there is no more reference to a bare action,
and that there is no need to introduce an UV regulator,
on account of the fact that the properties of ${\cal R}_k$
ensure that the r.h.s. is UV finite.
We shall now compute the beta function of $\tilde g$ by assuming that
the functional $\Gamma_k$ can be approximated by the form (\ref{action}), 
with $h_{\alpha\beta}$ fixed.
We thus neglect the effect of all higher derivative terms.
The resulting RG equation has almost exactly the same form as
(\ref{expansion}),
except for the appearance of a derivative of $\zeta$ on the r.h.s.,
which is due to the fact that the factor of $\zeta$ contained in ${\cal R}_k$ 
is now a renormalized, and therefore $k$--dependent, coupling:
$$
\dot{\Gamma}_{k}=
\frac{1}{2}\textrm{Tr}\,\frac{(\dot R_k+\eta R_k)\mathbf{1}}
{P_k(-D^{2})\mathbf{1}-\mathbf{M}}\ ,
$$
where $\eta=\dot\zeta/\zeta$. 
The relevant term in the trace is now
$$
\frac{1}{2}\textrm{Tr}\,\frac{\dot{R}_{k}+\eta R_{k}}{P_{k}^{2}}\mathbf{M}
= \frac{1}{2}c_{d}k^{d-2}\left(2+\frac{\eta}{\frac{d}{2}+1}\right)\int dx\sqrt{g}\, 
\partial_\mu\varphi^\alpha\partial^\mu\varphi^\beta R_{\alpha\beta}\,,
$$
whence we obtain
$$
\dot\zeta= 2c_{d}k^{d-2}\left(1+\frac{\eta}{d+2}\right)\frac{R}{D}\ .
$$
When this is solved for $\dot\zeta$ one obtains a rational function.
The beta function for the dimensionless coupling
$\tilde g$ is then
\begin{equation}
\label{exactbeta}
\beta=\frac{d-2}{2}\tilde{g}
-\frac{c_{d}\frac{R}{D}\tilde g^3}{1-2c_{d}\frac{R}{D(d+2)}\tilde g^2}\ .
\end{equation}
This beta function is our main result
\footnote{For $d>2$ its coefficients depend on the choice of cutoff $R_k$, 
but one can show that the qualitative properties of the beta function 
are the same for any cutoff.}.
When $R>0$ it has a FP at $\tilde g^2_*=\frac{1}{2}\frac{D(d^2-4)}{c_d dR}$.
Since the second term in the denominator of (\ref{exactbeta}) is positive,
the FP is always closer to the origin than at one loop.
The slope of the beta function at the FP is equal to $\beta'_*=-\frac{2d(d-2)}{d+2}<2-d$,
so it is steeper than at one loop (in particular $\nu=3/8$ in $d=4$).
Note that $\eta=d-2-2(\dot{\tilde g}/\tilde g)$,
so the anomalous dimension is equal to $d-2$ at any nontrivial FP.
Numerically, the results do not differ very much from one loop,
but since their derivation is not based on perturbation theory,
{\it their validity does not depend on the coupling being small}.

We conclude with some comments.
This work is at least partly motivated by the ongoing
search for a nonperturbative treatment of gravity 
along the lines of the ``asymptotic safety'' programme \cite{reviews}.
The NLSM has many features in common with gravity,
already at the kinematical level \cite{perbook},
and comparison between the two theories may be useful.
Also the structure of the dynamics is very similar:
except for the factor $\sqrt{\mathrm{det}g}$ and for the
different contractions of the indices,
the action (\ref{action}) for a group--valued NLSM
and the Hilbert action for gravity both have the structure
$\zeta\int (g^{-1}\partial g)^2$
where $g$ is either a $G$--valued scalar field or the metric,
and $\zeta$ has dimension mass$^{d-2}$.
The present work confirms that these analogies extend also to the
properties of the RG flow.
Existing results for higher derivative gravity \cite{higher,fofrall} 
suggest that the inclusion of higher terms in the NLSM will not spoil the FP. 
This will have to be checked.

Aside from being a possible toy model for gravity,
the NLSM has important applications to phenomenology.
The $SU(2)$ NLSM can be used as a low energy approximation to massless QCD
describing the dynamics of pions.
It is worth mentioning that in a fictional world where only massless pions existed,
an UV FP would unitarize the $\pi\pi\to\pi\pi$ scattering amplitude:
at tree level this amplitude grows like $g^2 E^2$,
where $E$ is some combination of external momenta,
but recalling that physics at the scale $k$ is described by the
action $\Gamma_k$ treated at tree level,
and identifying $k\approx E$, we see that in the FP regime the amplitude 
would tend to the constant $\tilde g_*$.
Unfortunately it is hard to see how this could be used in a realistic 
description of strong interactions, 
because at high enough energies one encounters many hadronic 
states that invalidate the simple NLSM description.

An ``asymptotically safe'' NLSM could be more useful in weak interaction physics.
In fact, the $SO(4)$ NLSM can be regarded as the strong coupling
limit of the scalar sector of the standard model.
Replacing the complex Higgs doublet by a $S^3$ NLSM results in a ``Higgsless'' theory.
Normally this is regarded only as an approximate description
valid below some cutoff of the order of the mass of the Higgs particle,
but if there is a FP, and assuming that there are no resonances,
then the Higgsless theory could hold up to much higher energies.
We plan to return to these issues elsewhere.

We should mention here that according to lattice calculations the triviality 
of $\phi^4$ theory in $d=4$ is expected to extend also to the corresponding NLSM
\cite{luscher,smit}.
It will be interesting to understand how our results fit with this expectation.
In this connection we observe that a nontrivial FP in the NLSM is not ruled out by a recent
investigation of the triviality issue using functional RG methods \cite{rosten}.
It may also be useful to repeat and improve the numerical simulations of \cite{bhanot}.

\medskip
\centerline{Acknowledgements}
We would like to thank J. Ambj\o rn, M. Reuter, and J. Smit for discussions.

\end{document}